\documentclass[sigconf]{acmart}

\usepackage[utf8]{inputenc}
\usepackage{natbib}
\usepackage{hyperref}
\usepackage{lineno} 
\usepackage{graphicx}
\usepackage{float}
\usepackage{enumitem}

\usepackage{booktabs}
\usepackage{colortbl}
\usepackage{multirow}
\usepackage{threeparttable}
\usepackage{subcaption}

\graphicspath{ {./figures/} }

\usepackage{tikz}
    \usetikzlibrary {positioning}
    \definecolor {processblue}{cmyk}{0.96,0,0,0}
    \usetikzlibrary{calc}
    \usetikzlibrary{shapes.geometric}
    \usetikzlibrary{backgrounds}
    \usepackage{varwidth}

\newcommand{\setItemSep}{\setlength\itemsep}
\newcommand{\tcolB}{\textcolor{blue}}
\newcommand{\tcolR}{\textcolor{red}}

\acmConference[CASCON'23]{33nd Annual International Conference on Computer Science and Software Engineering}{September 11-14, 2023}{
Las Vegas, NV, US}
\setcopyright{rightsretained} 
\acmDOI{}
\acmISBN{}

\begin{document}\sloppy

\title{Data Augmentation for Conflict and Duplicate Detection in Software Engineering Sentence Pairs}
\author{Garima Malik}
\affiliation{
  \institution{Toronto Metropolitan University}
  \city{Toronto}
  \state{Ontario}
  \country{Canada}
  }
\email{garima.malik@torontomu.ca}
\author{Mucahit Cevik}
\affiliation{
  \institution{Toronto Metropolitan University}
  \city{Toronto}
  \state{Ontario}
  \country{Canada}
  }
\email{mcevik@torontomu.ca}
\author{Ayşe Başar}
\affiliation{
  \institution{Toronto Metropolitan University}
  \city{Toronto}
  \state{Ontario}
  \country{Canada}
  }
\email{ayse.bener@torontomu.ca}


\begin{abstract}
This paper explores the use of text data augmentation techniques to enhance conflict and duplicate detection in software engineering tasks through sentence pair classification. 
The study adapts generic augmentation techniques such as shuffling, back translation, and paraphrasing and proposes new data augmentation techniques such as Noun-Verb Substitution, target-lemma replacement and Actor-Action Substitution for software requirement texts. A comprehensive empirical analysis is conducted on six software text datasets to identify conflicts and duplicates among sentence pairs. The results demonstrate that data augmentation techniques have a significant impact on the performance of all software pair text datasets. On the other hand, in cases where the datasets are relatively balanced, the use of augmentation techniques may result in a negative effect on the classification performance.
\end{abstract}


\settopmatter{printfolios=true}
\maketitle

\section{Introduction}
\label{intro}
Sentence pair text classification is an important topic of research in natural language processing (NLP), with several applications including Semantic Textual Similarity (STS), Paraphrase Identification (PI), Natural Language Inference (NLI), and Question Answering (QA) \citep{williams2017broad}.
The sentence pair text classification task takes two textual instances as input and the relationship between these texts is considered a class label.
For instance, the NLI problem consists of pairs of premise sentences as input, and the class label signifies whether the given premise pair is contradictory, neutral, or entailment.
\citet{malik2022transformer} adopt a similar approach to NLI and solve the requirement pair classification problem. 
They utilize sequential and cross-domain transfer learning approaches to categorize the relationships between the requirement pairs as conflict, duplicate, and neutral.
They note that the scarcity of labeled textual data is one of the major challenges in using transfer learning for various NLP tasks in the software engineering domain.
Similarly, \citet{he2020duplicate} studies the duplicate bug identification problem from textual bug reports using pair representation in bug tracking management systems. The duplicate text detection problem is also highly relevant to social collaborative platforms for software development such as Stack Overflow and Github, where the continuous expansion of duplicate questions in the databases poses significant challenges~\citep{silva2018duplicate}.

It is common in practice for software systems to have a broad user base, which makes it possible for multiple users to generate conflicting or duplicate requirements. Likewise, duplicate bugs and duplicate questions on social media platforms can also be generated easily.
A significant amount of time and effort might be saved on software text analysis by automatically identifying the conflicts, duplicates, and neutral texts~\citep{xie2018detecting}. On the other hand, the labeled training data must be large enough to create accurate text classifiers such that the model can generalize to unseen data. The lack of labeled data for model training is a frequently encountered problem in the software engineering domain.
Accordingly, it is important to take advantage of recent progress in data augmentation (DA) techniques for text classification to improve the performance of sentence pair classification models over software texts and further advance the state-of-the-art for conflict and duplicate identification in this domain.

Conflicting requirement specifications for software development and redundant or duplicate bug reports in the task management systems might lead to various inefficiencies in software systems. 
As a result, it is important to explore how NLP techniques can be used to enhance the automated processing of software texts.
Such automation offers chances to streamline the software development processes and minimize potential issues that might occur when interpreting the software requirements, bugs, and programming-related social media posts.
For instance, the software development teams can save significant time and effort by using an automated method for labeling any given requirement as conflicting, duplicate, or neutral to any other requirement in the database. High-accuracy sentence pair text classification models can be thought of as building blocks for such automation.

\paragraph{\textbf{Research Contribution}}
The main contributions of our study can be summarized as follows:
\begin{itemize}\setItemSep{0.4em}
    \item We evaluate various data augmentation techniques and customize these approaches for sentence pair classification tasks. To the best of our knowledge, this constitutes the first study that provides a systematic evaluation of text data augmentation strategies for sentence pair classification tasks in software requirement engineering.

    \item We propose three novel data augmentation strategies, noun-verb substitution using WordNet, target replacement using WordNet lemmas, and actor-action substitution using Word2vec, which are well-suited for software sentence pairs. 
    
    \item We conduct a comprehensive numerical study with six different software engineering datasets and we demonstrate that the performance of transformer models can be enhanced by employing data augmentation strategies. Additionally, these DA techniques are applied across paired datasets in seven distinct combinations, and the effectiveness of each configuration is thoroughly evaluated.
\end{itemize}

\paragraph{\textbf{Structure of the paper}} 
The rest of the paper is organized as follows. Section~\ref{sec:literature} provides an overview of various data augmentation methods applied to NLP tasks. 
Section~\ref{sec:methods} provides a summary of the data augmentation methodologies, dataset descriptions, and a detailed discussion of the experimental design. 
In Section~\ref{sec:results}, we present the results of our numerical study, which includes a performance comparison of various data augmentation strategies.
Lastly, Section~\ref{sec:conc} provides concluding remarks and future research directions.
\section{Literature Review} \label{sec:literature}
The term ``data augmentation'' refers to a wide variety of techniques that have their roots in the field of computer vision.
For instance, rotations and alterations to the RGB channel are found to be effective transformations for images.
Speech recognition uses processes that alter sound or speed, much like computer vision \citep{bayer2021survey}. In contrast, developing such uniform procedures for textual data modifications can be challenging, as it might be difficult to retain the labeling quality and syntactic meaning for NLP tasks~\citep{wei2019eda,karimi2021aeda,shorten2021text}.

\citet{wei2019eda} presented EDA (Easy Data Augmentation): simple strategies for data augmentation to improve performance on text classification tasks.
The four straightforward yet effective procedures in EDA are synonym substitution, random insertion, random swap, and random deletion.
They demonstrated that EDA enhances performance for both convolutional and recurrent neural networks on five text classification tasks.
\citet{karimi2021aeda} proposed AEDA (An Easier Data Augmentation) as a simpler data augmentation strategy for text classification.
Only punctuation mark insertions into the original text are included by AEDA and it maintains the word order.
\citet{xiang2020lexical} employed part-of-speech (POS) focused lexical substitution for data augmentation (PLSDA) to improve deep learning model performance and prediction capabilities.
PLSDA finds words by using part-of-speech information and then uses various augmentation techniques to locate semantically relevant replacements to produce new instances for training.
PLSDA was applied to four deep learning models and it was shown to enhance their accuracy by an average of 1.3\%.

\citet{zhang2015character} employed a text augmentation technique that replaces words or phrases with synonym sets obtained from WordNet English Thesaurus. 
\citet{huang2019glossbert} posited that the gloss forms found in the WordNet database could be employed to exhibit diverse semantics and variations in the text. Therefore, they harnessed these gloss forms for the task of word sense disambiguation (WSD). \citet{marivate2020improving} highlighted the effectiveness of Word2vec-based data augmentation techniques for textual data. They introduced a Synonym augmentation strategy, which utilizes dictionaries and distributed word representations to identify semantically similar words. In addition to this, they implemented Round-trip translation (RTT), further enhancing their augmentation methods. \citet{perccin2022combining} proposed a novel DA method that combined the WordNet and word embeddings (GloVe) for legal text analysis.

\citet{chen2020local} implemented a local additivity-based data augmentation method, in which augmented text instances are produced by interpolating the hidden representations of the tokens with tokens from either the same text instance or tokens from a different text instance.
The translation data augmentation model introduced by \citet{fadaee2017data} used the back translation technique in which English text instances are first translated into German and then back to English while the semantics of the text samples are maintained. \citet{kumar2020data} investigated various transformer-based pre-trained models for conditional data augmentation, including auto-regressive models (GPT-2), auto-encoder models (BERT), and Seq2Seq models (BART).
They demonstrated that a straightforward yet successful method for preparing the pre-trained models for data augmentation is to prefix the class labels to text sequences. Additionally, the pre-trained Seq2Seq model was shown to beat alternative data augmentation techniques in a low-resource environment on three classification benchmarks.

To the best of our understanding, current literature on DA approaches for textual data primarily focuses on generic NLP tasks such as text classification, word sense disambiguation and sentiment analysis. We build upon these traditional DA methods and introduce novel strategies specifically designed for software engineering datasets that utilize paired inputs, thus filling a unique gap in the existing research.

\section{Methodology}\label{sec:methods}
In this section, we first describe the datasets employed in our analysis, then we elaborate on various data augmentation techniques used in the experiments. 
Lastly, we provide details on the experimental setup.
\subsection{Datasets}\label{subsec:data}
We consider four (WorldVista, UAV, PURE, and OPENCOSS) open-source software requirement datasets where requirement pairs are labeled as conflict and Neutral. 
We also consider two software platform datasets (Stack Overflow and Bugzilla ) which consist of paired texts in the form of social media posts and bug reports extracted from the web and online repositories. 
These paired bugs and posts are labeled as neutral and duplicate. 
Table~\ref{tab:data_dist} provides the summary information for all the datasets.
Table~\ref{tab:sample_data} provides sample data instances that demonstrate the general structure and format of the requirement texts, bug reports, and social media posts.
\setlength{\tabcolsep}{4.5pt}
\renewcommand{\arraystretch}{1.1}
\begin{table}[!ht]
\centering
    \caption{Class distribution of software requirements and software platform datasets}
    \label{tab:data_dist}
    \resizebox{0.335\textwidth}{!}{
    \begin{tabular}{lrr}
\toprule
\textbf{Dataset}       & \textbf{\# Neutral } & \textbf{\# Conflict} \\ 
\midrule
UAV           & 6,652              & 18                             \\
WorldVista    & 10,843              & 35                             \\
PURE          & 2,191               & 20                             \\
OPENCOSS      & 6,776                & 10                             \\
\midrule
 & \textbf{\# Neutral } & \textbf{\# Duplicate} \\
\midrule
StackOverflow & 5,000                & 90                            \\
BugZilla      & 4,000                & 90                            \\ \bottomrule
\end{tabular}
    }
\end{table}

\setlength{\tabcolsep}{4.5pt}
\renewcommand{\arraystretch}{1.1}
\begin{table*}[!ht]
\centering
    \caption{Sample paired text instances from software engineering datasets. 
    The `Label' column indicates the relationship between Text 1 and Text 2. The highlighted words within the text serve to indicate the reasoning behind the application of a specific label.}
    \resizebox{0.99\linewidth}{!}{
    \begin{tabular}{l p{0.35\textwidth} p{0.35\textwidth} p{0.08\textwidth} }
\toprule
\textbf{Dataset}       & \textbf{Text 1} & \textbf{Text 2} & \textbf{Label} \\ \midrule
UAV           & The \tcolB{\_InternalSimulator\_} shall provide the exact state of the battery. &  The \tcolB{\_VehicleCore\_} shall support up to three virtual UAVs& Neutral     \\
StackOverflow & \tcolB{Accessing} a \tcolB{variable} within a nested function of the same class I am trying to access variable within function\_3 how should I go about doing this? & \tcolB{Merging} two \tcolB{dataframes} and updating count and date I have this dataframe: And this dataframe: How can I merge them by the number & Neutral     \\
\midrule
WorldVista    & The \tcolB{system's pilot program} shall use a smart card to \tcolB{digitally sign} medication orders & The \tcolB{system's pilot program} shall require a \tcolB{handwritten signature} for medication orders & Conflict    \\
PURE          & The \tcolB{supervisor’s interface} \tcolR{shall} display the available thermostats and the individual current temperature settings & The \tcolB{supervisor’s interface} \tcolR{must} display the  thermostats and the individual temperature settings & Conflict    \\
\midrule
Bugzilla      & Implement \tcolB{drag and drop }of attached messages to mail folders & Cannot \tcolB{drag and drop} an attached email onto a folder I have a message with another message as an attachment & Duplicate \\
Stack Overflow & Why is my \tcolB{recursive function} returning a None value & Why does my \tcolB{recursive function} return None? & Duplicate \\
 \bottomrule
\end{tabular}
    }
    \label{tab:sample_data}
\end{table*}

Below, we briefly describe the datasets used in our analysis.
\begin{itemize}\setItemSep{0.6em}
    \item \textbf{UAV}: The University of Notre Dame released the UAV (Unmanned Aerial Vehicle) dataset, which contains all the required specifications that explain the operations of the UAV control system automatically. The requirement syntax adheres to the EARS (Easy Approach to Requirements Syntax) template \citep{mavin2009easy}. There are only a few conflicts present in this dataset, which result in 18 conflicts and 6,652 neutral pairs.

    \item \textbf{WorldVista}: The WorldVista\footnote[1]{https://worldvista.org/Documentation} dataset covers the software requirements for a health management system. 
    The patient's information, including hospital admission and discharge processes, is recorded by this software system. 
    The format of the requirements is very basic, expressed in natural language with primitive health care terminologies. 
    This dataset consists of 35 conflict and 10,843 neutral pairs.
    
    \item \textbf{PURE}: PURE (Public Requirements dataset) is a publicly accessible corpus that incorporates 79 SRS documents collected from the open-source software projects \citep{pure}. 
    In particular, we chose two SRS documents from this collection named THEMAS (Thermodynamic System) and Mashbot (web interface for social networks). 
    The initial requirement structures were intricate and divided into paragraphs. 
    The requirements were preprocessed into less complex forms.  
    This dataset contains 20 conflict and 2,191 neutral pairs.

     \item \textbf{OPENCOSS}: The OPENCOSS\footnote{http://www.opencoss-project.eu} project focuses on the development of an Open Platform for the Evolutionary Certification of Safety-critical Systems, with particular emphasis on the railway and automotive sectors. This dataset poses a distinct challenge owing to its narrow scope, having 10 conflict pairs compared to the substantially larger number of 6,776 neutral pairs.
    
    \item \textbf{Stack Overflow}: Stack Overflow website specializes in solving software and programming-related issues. 
    The website contains duplicate questions despite efforts to prevent asking questions that have already been addressed. 
    We extracted neutral and duplicate pairs of questions related to various software programming languages. 
    To maintain consistency with other datasets, we primarily experimented with 5,000 neutral pairs and 90 duplicate pairs.
    
    \item \textbf{Bugzilla}: Bugzilla is one of the widely used bug tracking systems where users create bug reports for a given software. However, because of the lack of coordination and distribution in this process, numerous users may file bug reports describing the same issue, i.e., duplicate bug reports. 
    For our primary experimentation, we extracted 4,000 neutral pairs and 90 duplicate pairs.
    
\end{itemize}

\subsection{Data Augmentation Methods}\label{subsec:DA_methods}
\begin{figure*}[!ht]
    \centering
    \includegraphics[width=0.905\textwidth]{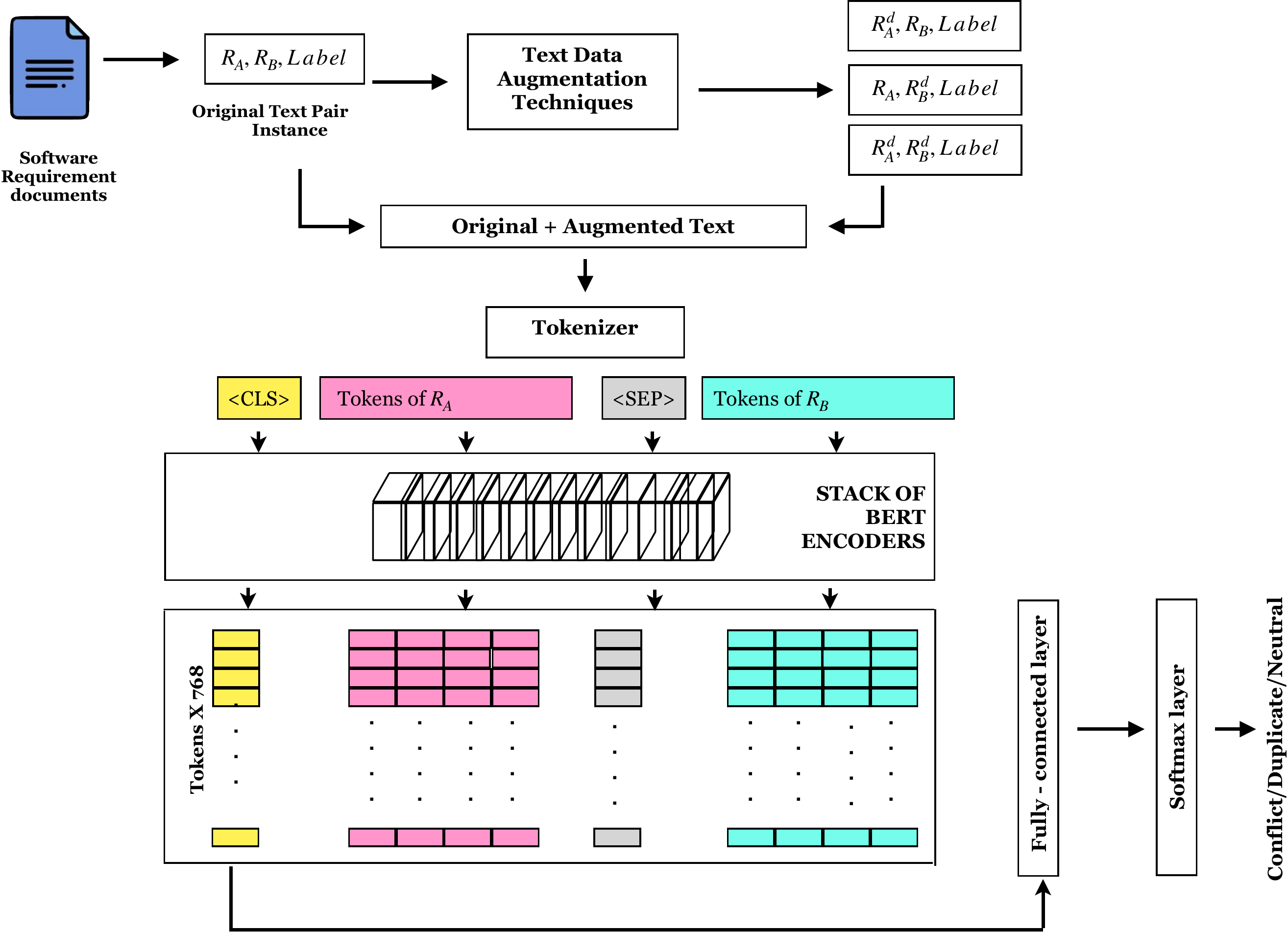}
     \caption{Working flow of experimentation with data augmentation strategies for software sentence pair classification}
     \label{fig:flow_chart}
\end{figure*}

The sentence pair classification task utilizes text data augmentation strategies, as shown in Figure~\ref{fig:flow_chart}.
To augment the requirement texts, we have implemented a procedure that involves a set of datasets $D = \{D_1, D_2, D_3, D_4, D_5, D_6\}$, where each dataset $D_i$ contains pairs of requirements and their corresponding labels as $(R_A, R_B,Label)$. We apply a data augmentation function $DA(R_A)$ to each requirement $R_A$. We have implemented the following cases:
\begin{itemize}\setItemSep{0.5em}
    \item \textbf{Case I}: We apply the DA procedure to the first text of each pair in each dataset $D_i \in D$. For each pair $(R_A, R_B,Label) \in D_i$, we transform it into $(DA(R_A), R_B, Label)$.
    \item \textbf{Case II}: We apply the DA procedure to the second text of each pair in each dataset $D_i \in D$. For each pair $(R_A, R_B,Label) \in D_i$, we transform it into $(R_A, DA(R_B), Label)$.
    \item \textbf{Case III}: We apply the DA procedure to both texts of each pair in each dataset $D_i \in D$. For each pair $(R_A, R_B,Label) \in D_i$, we transform it into $(DA(R_A), DA(R_B),Label)$.
\end{itemize}

Furthermore, we combine these cases in the following ways:
Case II+III, which involves applying the DA procedure to the second text of each pair and both texts of each pair in each dataset.
Case I+III, which involves applying the DA procedure to the first text and both texts of each pair in each dataset.
Case I+II, which involves applying the DA procedure to the first and second texts of each pair in each dataset.
Case I+II+III, which involves applying the DA procedure to all texts of each pair in each dataset. The DA procedure utilized in each case can be any of the data augmentation techniques described in the next section.
\subsubsection{Traditional Data Augmentation Methods}
Traditional techniques for data augmentation in NLP include synonym replacement, random insertion of words, random deletion of words, and random swapping of words \cite{bayer2021survey}. 
In our analysis, shuffling, back translation, and paraphrasing were considered  generic DA techniques because they do not involve any specific customization for the software requirement engineering and code data. 
These techniques are known to preserve the structure and nature of the original text.

\begin{itemize}\setItemSep{0.6em}
    \item \textbf{Shuffling}: Shuffling is a DA technique that randomly shuffles the structure of input text to create augmented text. 
    Random swaps of words in the text can also be considered as a DA technique for text classification tasks \citep{wei2019eda}. 
    That is, some random words can be selected from the input sentence and the position of words can be changed to generate the augmented text. 
    This technique is expected to perform well with longer input texts present in the Bugzilla and Stack Overflow datasets.
    
    \item \textbf{Back Translation}: Back Translation is an approach to creating textual data with the help of language translation models \citep{howard2018universal}. 
    Textual data in the form of words, phrases, sentences, and documents are translated into another language using forward translation; back translation is the translation of the same text back into the original language \citep{bayer2021survey}. 
    The use of the back translation strategy as a DA technique can be justified by the fact that translations of texts frequently vary due to the complexity of natural language which creates a wide range of potential terminology or sentence structures \citep{edunov2018understanding}. 
    This method preserves the original text labels while translating the text and only the stylistic elements are altered, depending on the source text \citep{bayer2021survey}. 
    The highly relevant paraphrasing capabilities of back translation models make it a suitable technique to augment the software-paired text datasets.
    As the software text structure follows certain templates, it is very important to preserve the structure and format of the software text. 
    For our experiments, we tried different languages for a back translation, and the German language provided appropriate augmented sentences. 
    To implement back translation, we use nlpaug\footnote[2]{https://github.com/makcedward/nlpaug} library and pre-trained transformer model from fairseq\footnote[3]{https://github.com/facebookresearch/fairseq}  (provided by Facebook AI Research). 
    Each training instance is translated from the source input (English) to the target input (German) and then translated back to the original language.
    
    \item \textbf{Paraphrasing}: The task of paraphrasing is similar to document summarization in NLP. In paraphrasing, the transformer-based sequence model generates a textual output that implies the same meaning as the given input text. 
    As paraphrased text context is similar to the original text, we expect to generate meaningful augmented requirements and bug reports from the paraphrasing technique. 
    To implement paraphrasing, we use PEGASUS (Pre-training with Extracted Gap-sentences for Abstractive Summarization) framework \citep{zhang2020pegasus}. 
    PEGASUS is a standard encoder-decoder transformer model which implements gap sentence generation (GSG) and mask language modeling (MLM) simultaneously. 
    For our experiments, we generate ten paraphrased sentences for each input instance and add them to the training set.
\end{itemize}
\subsubsection{Proposed Data Augmentation Approaches}
We propose three custom techniques for software sentence pair augmentation, which are based on the characteristics of the input data i.e., software requirements, software codes, and bug reports.
\begin{itemize}\setItemSep{0.6em}
     \item \textbf{Noun-Verb Substitution using WordNet Synsets (NV-WNS)}: Synonym replacement is one of the primitive data augmentation techniques for text classification tasks in NLP. 
     In synonym replacement, random words from the input instance are replaced by the respective synonyms present in the thesaurus. 
     Instead of using synonym replacement with random words in software text, we create a customized version of synonym replacement where we only replace the noun and verb present in the text with their synonyms extracted from WordNet\footnote[4]{https://wordnet.princeton.edu/} (lexical database). 
     Nouns and verbs are key elements of requirements as they convey vital information about the requirement. 
     In the requirement statement, \textit{``The \_VehicleCore\_ shall support up to three virtual UAV's''}, both the \textit{\_VehicleCore\_} and \textit{UAV's} are examples of nouns, while \textit{support} is an example of a verb. 
     These nouns and verbs together specify what is required of the `VehicleCore', and what action it needs to take. 
     Using part of speech tagging from the NLTK library, we first identify the noun and verb for each input text. 
     Then, to create the augmented text, we extract the synonym sets from WordNet and replace the noun and verb in the original text with the synonyms.
     
     \item \textbf{Actor-Action Substitution using word2vec (AA-W2V)}: The two important entities in software requirement texts are the actor and the action. 
     For instance, consider the requirement: \textit{``The aviary shall fly with the speed of $20m/s^2$''}. Here, \textit{`aviary'} is the actor and \textit{`fly'} is the action. 
     We use these entities (actor and action) as potential candidates for synonym replacement in the input text. 
     Firstly, we identify the actor and action from each software requirement using a software-specific named entity recognition model \citep{malik2022identifying}. 
     Then, we take the words that are the closest in meaning to the actor and action and extract them from word2vec embeddings. 
     We replace the actor and action in software requirements from a similar word set and generate augmented sentences.
     NV-WN techniques also use word substitution, but the distinction between NV-WN and AA-W2V lies in the fact that AA-W2V assumes only one actor and one action in the requirement statement, whereas NV-WN can handle multiple nouns and verbs in the same statement. 
     Therefore, while AA-W2V is limited to simple requirements with one actor and one action, NV-WN can handle more complex requirements.
     The AA-W2V technique was not applied to the Bugzilla and Stack Overflow datasets because these datasets are not related to software requirements. 
     That is, the AA-W2V technique is designed specifically for software requirement datasets where the actor and action can be identified.

     \item \textbf{Target Replacement using WordNet Lemma (T-WNL)}: The T-WNL technique is a methodology based on the approach described in the research article by \citep{huang2019glossbert}, which proposed the use of GlossBERT for Word Sense Disambiguation (WSD) tasks. In the GlossBERT technique, the target words in the dataset are replaced with their corresponding Gloss form in WordNet. We examine different aspects of WordNet and find that synsets, lemmas, gloss examples, and grammatical relations are provided for each word. After analyzing our dataset, we determined that lemmas are more suitable for our purposes.
     
     Primarily, this technique involves the extraction of six pivotal elements from the software text, namely the actor, action, object, property, metric, and operator. These elements are subsequently deemed as target words within the text. Following the extraction, each target word is cross-referenced with the WordNet lexical database to identify its corresponding lemma form. The original target words in the software text are then replaced with these lemma forms, resulting in an augmented requirement. It is crucial to note that the replacement process is conducted individually for each lemma, leading to the creation of unique augmented requirements. For instance, consider the software requirement: \textit{``The UAV shall fully charge in less than 3 hours"}. In this context, \textit{`UAV'} is identified as the actor, \textit{`charge'} as the action, \textit{`less than'} as the operator, and \textit{`3 hours'} as the metric. These target elements are precisely identified utilizing a software-specific NER model as described in \citep{malik2022identifying}. By implementing the T-WNL technique, we systematically alter these targets with their respective lemma forms, thereby generating a diverse set of augmented software requirements.

     \item \textbf{Combined DA}: In this technique, we combine the instances generated from all the text data augmentation techniques and sample the instances equivalent to the neutral class instances in the training set. This helps us to combine the advantages of all the text data augmentation techniques and provides variability in the augmented set. By combining the instances generated from all the data augmentation techniques, we create a larger and more diverse training set.
\end{itemize}

\subsection{BERT Model Training}
Once the data augmentation techniques have been applied to the training set, the augmented instances are added to the training set and fed into the BERT tokenizer, as shown in Figure~\ref{fig:flow_chart}. The BERT tokenizer adds a special <CLS> token at the beginning of the first software text, $R_A$, and a <SEP> token at the beginning of the second software text, $R_B$. 
These special tokens are used to indicate the beginning and separation of the input texts. 
The tokenized texts are then fed into the BERT encoder, which generates a 1x768 representation for each token in the input sequence. 
These representations capture the contextual information of each token and are aggregated to generate a final representation of the input sequence. 
The final representation of <CLS> token is then passed through a softmax function to obtain the class probabilities for the input instance.

\subsection{Experimental Setup}
We use a pre-trained transformer model from huggingface\footnote [5]{https://huggingface.co/docs/transformers/index} library for software pair text classification task. 
For all the datasets, we use \textit{bert-base-uncased-MNLI}\footnote[6]{https://huggingface.co/textattack/bert-base-uncased-MNLI} transformer checkpoint. This transformer checkpoint is identified based on our preliminary analysis. We apply various DA techniques such as shuffling, back translation, paraphrasing, Noun and Verb replacement with WordNet, and Actor and Action replacement with pre-trained word2vec\footnote[7]{https://huggingface.co/fse/word2vec-google-news-300} embedding with dimension value of 300. 

Table~\ref{tab:hyper} provides the details of the transformer model checkpoint and hyperparameters used in our experiments. We employ an adaptive learning rate scheduler that modifies the learning rate during the hyperparameter tuning process of the transformer models.
We set up various configurations of training epochs (2,10) and training batch sizes (8,128). 
Contrary to usual deep learning practice, we find that model checkpoints perform better with smaller batch sizes and fewer number epochs. We use early stopping callback to track the validation loss and get the fine-tuned model checkpoint for testing.
\begin{table}[!ht]
\centering
    \caption{\textit{bert-base-uncased-MNLI} (Layer = 12, Hidden units = 768, Total params = 340M) transformer model hyperparameter values for all the datasets}
    \resizebox{\columnwidth}{!}{
    \begin{tabular}{p{2 cm}|rrrr}
\toprule
\textbf{Datasets} & \textbf{Batch size} & \textbf{Epochs} & \textbf{Max. length} & \textbf{Metrics} \\
\midrule
WorldVista, UAV PURE, \& OPENCOSS  & 32         & 10     & 128         & Conflict-F1    \\
\midrule
StackOverflow \& BugZilla  & 8          & 5      & 512         & Duplicate-F1 \\
\bottomrule
\end{tabular}
    }
    \label{tab:hyper}
\end{table}

\section{Results} \label{sec:results}
In this section, we present experimental results comparing the performance of various DA techniques on the sentence pair software-specific datasets. As the evaluation criteria, we consider standard classification metrics such as the macro version of the F1-score and to highlight the performance improvement in the minority class i.e., conflict and duplicate class in the datasets, we also report conflict class F1-score and duplicate class F1-score. 

\subsection{Performance Comparison} 
This section presents an analysis of the numerical results obtained from two main perspectives: firstly, we assess the performance impact of various DA configurations on software requirement pair datasets, followed by an evaluation of these methods on software platform duplicate post-classification datasets.

Table~\ref{tab:conflict_results} reports the enhancements in conflict class F1-scores relative to the `No Augmentation' baseline across all requirement pair datasets. A systematic analysis of our results reveals that the application of DA techniques facilitates significant performance improvements for WorldVista, UAV, and PURE datasets. However, we observe a contrary trend with OPENCOSS, where the DA techniques seem to lower the performance.

\begin{table*}[!ht]
\centering
\caption{Empirical analysis of DA techniques on conflict class F1-score for \textit{software requirement datasets}. The reported metrics is averaged over three folds and presented as ``mean ± standard deviation''. The table presents a comparative overview against a ``No Augmentation'' baseline, with absolute performance changes color-coded: blue for positive and red for negative shifts. Grey color highlights indicate the most effective augmentation method per requirement pair dataset.}
\label{tab:conflict_results}
    \subfloat[WorldVista (No Augmentation conflict-F1: 0.817 $\pm$ 0.087) \label{tab:w_v_full_report}]{
\resizebox{0.985\textwidth}{!}{
\begin{tabular}{l|rrrrrrr}
    \toprule
        & I   & II  & III  & I+II & I+III & II+III & I+II+III   \\
    \midrule
Shuffling & 0.858 $\pm$ 0.079 (\tcolB{+0.04})  & 0.846 $\pm$ 0.025 (\tcolB{+0.02}) & 0.904 $\pm$ 0.040 (\tcolB{+0.08}) & 0.841 $\pm$ 0.093 (\tcolB{+0.02})  & 0.880 $\pm$ 0.108 (\tcolB{+0.06}) & \cellcolor{gray!20}\textbf{0.908 $\pm$ 0.006 (\tcolB{+0.09})} & 0.871 $\pm$ 0.064 (\tcolB{+0.05})\\

Back Translation & 0.807 $\pm$ 0.073 (\tcolR{- 0.01}) & 0.879 $\pm$ 0.014 (\tcolB{+0.06}) & 0.866 $\pm$ 0.066 (\tcolB{+0.04}) & 0.846 $\pm$ 0.093 (\tcolB{+0.02}) & 0.864 $\pm$ 0.093 (\tcolB{+0.04})  & 0.838 $\pm$ 0.083 (\tcolB{+0.02})  & \cellcolor{gray!20}\textbf{0.892 $\pm$ 0.016 (\tcolB{+0.07})} \\

Paraphrasing & 0.891 $\pm$ 0.066 (\tcolB{+0.07}) & 0.890 $\pm$ 0.090 (\tcolB{+0.07}) & 0.847 $\pm$ 0.080 (\tcolB{+0.03}) & 0.856 $\pm$ 0.089 (\tcolB{+0.03}) & 0.890 $\pm$ 0.058 (\tcolB{+0.07}) &  0.858 $\pm$ 0.100 (\tcolB{+0.04}) &  \cellcolor{gray!20}\textbf{0.902 $\pm$ 0.046 (\tcolB{+0.08})}\\
\midrule
NV-WNS &0.788 $\pm$ 0.142 (\tcolR{- 0.02}) &0.767 $\pm$ 0.092 (\tcolR{- 0.05}) & 0.824 $\pm$ 0.091 (\tcolB{$\sim$0.00}) & \cellcolor{gray!20}\textbf{0.885 $\pm$ 0.049 (\tcolB{+0.06})} &0.812 $\pm$ 0.050 (\tcolR{$\sim$0.00}) & 0.841 $\pm$ 0.082 (\tcolB{+0.02}) & 0.754 $\pm$ 0.106 (\tcolR{- 0.06}) \\

AA-W2V & \cellcolor{gray!20}\textbf{0.886 $\pm$ 0.049 (\tcolB{+0.06})}  & 0.813 $\pm$ 0.079 (\tcolR{$\sim$0.00}) & 0.844 $\pm$ 0.079 (\tcolB{+0.02}) & 0.883 $\pm$ 0.036 (\tcolB{+0.06})& 0.849 $\pm$ 0.057 (\tcolB{+0.03}) & 0.868 $\pm$ 0.087 (\tcolB{+0.05})& 0.879 $\pm$ 0.072 (\tcolB{+0.06}) \\

T-WNL & 0.828 $\pm$ 0.073 (\tcolB{+0.01}) & 0.853 $\pm$ 0.093 (\tcolB{+0.03}) & 0.821 $\pm$ 0.119 (\tcolB{$\sim$0.00}) & 0.866 $\pm$ 0.059 (\tcolB{+0.04}) & 0.841 $\pm$ 0.125 (\tcolB{+0.02}) & 0.857 $\pm$ 0.034 (\tcolB{+0.04}) & \cellcolor{gray!20}\textbf{0.875 $\pm$ 0.040 (\tcolB{+0.05})} \\
Combined DA & 0.839 $\pm$ 0.043 (\tcolB{+0.02}) & 0.846 $\pm$ 0.062 (\tcolB{+0.02}) & \cellcolor{gray!20}\textbf{0.895 $\pm$ 0.056 (\tcolB{+0.07})} & 0.839 $\pm$ 0.043 (\tcolB{+0.02}) & 0.840 $\pm$ 0.040 (\tcolB{+0.02}) & 0.854 $\pm$ 0.084 (\tcolB{+0.03}) & 0.806 $\pm$ 0.080 (\tcolR{- 0.01})\\

\bottomrule
\end{tabular}
        
    }} \\*[0.4em]
    \subfloat[UAV (No Augmentation conflict-F1: 0.698 $\pm$ 0.071) \label{tab:uav_full_report}] {
    \resizebox{0.985\textwidth}{!}{
    \begin{tabular}{l|rrrrrrr}
    \toprule
        &I   & II  & III  & I+II & I+III & II+III & I+II+III   \\
    \midrule
Shuffling & 0.811 $\pm$ 0.015 (\tcolB{+0.11})& \cellcolor{gray!15}\textbf{0.841 $\pm$ 0.063 (\tcolB{+0.14})}& 0.770 $\pm$ 0.045 (\tcolB{+0.07})& 0.729 $\pm$ 0.139 (\tcolB{+0.03}) & 0.774 $\pm$ 0.073 (\tcolB{+0.07})& 0.774 $\pm$ 0.133 (\tcolB{+0.07}) & 0.818 $\pm$ 0.080 (\tcolB{+0.12}) \\

Back Translation & 0.823 $\pm$ 0.074  (\tcolB{+0.12})& 0.744 $\pm$ 0.103  (\tcolB{+0.04})& \cellcolor{gray!15}\textbf{0.832 $\pm$ 0.117 (\tcolB{+0.13})}& 0.730 $\pm$ 0.089 (\tcolB{+0.03})& 0.774 $\pm$ 0.079 (\tcolB{+0.07})& 0.726 $\pm$ 0.055 (\tcolB{+0.02})& 0.747 $\pm$ 0.114 (\tcolB{+0.04}) \\

Paraphrasing & 0.671 $\pm$ 0.160 (\tcolR{- 0.02})& 0.675 $\pm$ 0.127 (\tcolR{- 0.02})& 0.755 $\pm$ 0.062 (\tcolB{+0.05})& 0.541 $\pm$ 0.058 (\tcolR{- 0.15}) & 0.709 $\pm$ 0.082  (\tcolB{+0.01})& \cellcolor{gray!15}\textbf{0.836 $\pm$ 0.051 (\tcolB{+0.13})}& 0.631 $\pm$ 0.096 (\tcolR{- 0.06})\\

\midrule

NV-WNS &0.719 $\pm$ 0.156 (\tcolB{+0.02})& 0.667 $\pm$ 0.135 (\tcolR{- 0.03})& 0.548 $\pm$ 0.040 (\tcolR{- 0.15})& 0.631 $\pm$ 0.096 (\tcolR{- 0.06})& 0.810 $\pm$ 0.104 (\tcolB{+0.11})& \cellcolor{gray!15}\textbf{0.872 $\pm$ 0.051} (\tcolB{+0.17})& 0.774 $\pm$ 0.079 (\tcolB{+0.07})\\

AA-W2V &\cellcolor{gray!15}\textbf{0.828 $\pm$ 0.114 (\tcolB{+0.13})}& 0.794 $\pm$ 0.053 (\tcolB{+0.09}) & 0.775 $\pm$ 0.034 (\tcolB{+0.07})& 0.722 $\pm$ 0.078 (\tcolB{+0.02}) & 0.747 $\pm$ 0.114 (\tcolB{+0.04})& 0.730 $\pm$ 0.089 (\tcolB{+0.03})& 0.791 $\pm$ 0.099 (\tcolB{+0.09})\\

T-WNL & 0.766 $\pm$ 0.072 (\tcolB{+0.06})& 0.759 $\pm$ 0.030 (\tcolB{+0.06})& 0.847 $\pm$ 0.045 (\tcolB{+0.14})& 0.812 $\pm$ 0.085 (\tcolB{+0.11})& \cellcolor{gray!15}\textbf{0.877 $\pm$ 0.055 (\tcolB{+0.17})}& \cellcolor{gray!15}\textbf{0.877 $\pm$ 0.087 (\tcolB{+0.17})}& 0.853 $\pm$ 0.089 (\tcolB{+0.15})\\

Combined DA & 0.756 $\pm$ 0.068 (\tcolB{+0.06})& 0.883 $\pm$ 0.114 (\tcolB{+0.18})& 0.812 $\pm$ 0.074 (\tcolB{+0.11})& 0.782 $\pm$ 0.122 (\tcolB{+0.08}) & 0.841 $\pm$ 0.058 (\tcolB{+0.14})& 0.907 $\pm$ 0.082  (\tcolB{+0.20})& \cellcolor{gray!15}\textbf{0.914 $\pm$ 0.068  (\tcolB{+0.21})} \\

\bottomrule
\end{tabular}
        
    }} \\*[0.4em]
    \subfloat[PURE (No Augmentation conflict-F1: 0.841 $\pm$ 0.011)\label{tab:pure_full_report}] {

\resizebox{0.985\textwidth}{!}{
\begin{tabular}{l|rrrrrrr}
    \toprule
        &I   & II  & III  & I+II & I+III & II+III & I+II+III   \\
    \midrule
Shuffling & 0.832 $\pm$ 0.119 (\tcolR{$\sim$0.00})& 0.823 $\pm$ 0.074 (\tcolR{- 0.01})& 0.853 $\pm$ 0.112 (\tcolB{+0.01})& 0.853 $\pm$ 0.089  (\tcolB{+0.01}) & 0.835 $\pm$ 0.081 (\tcolR{- 0.00})& \cellcolor{gray!15}\textbf{0.888 $\pm$ 0.039 (\tcolB{+0.04})} & 0.806 $\pm$ 0.084 (\tcolR{- 0.03})\\

Back Translation & 0.836 $\pm$ 0.030 (\tcolR{$\sim$0.00}) & 0.780 $\pm$ 0.131 (\tcolR{- 0.06}) & \cellcolor{gray!15}\textbf{0.918 $\pm$ 0.068 (\tcolB{+0.07})} & 0.842 $\pm$ 0.115 (\tcolB{$\sim$0.00})& 0.867 $\pm$ 0.097 (\tcolB{+0.02})& 0.857 $\pm$ 0.092 (\tcolB{+0.01})& 0.833 $\pm$ 0.000 (\tcolR{$\sim$0.00})\\

Paraphrasing & 0.923 $\pm$ 0.000 (\tcolB{+0.08})& 0.776 $\pm$ 0.043 (\tcolR{- 0.06}) & 0.893 $\pm$ 0.042 (\tcolB{+0.05})& 0.861 $\pm$ 0.111 (\tcolB{+0.02})& \cellcolor{gray!15}\textbf{0.926 $\pm$ 0.058 (\tcolB{+0.08})}& 0.835 $\pm$ 0.081 (\tcolR{$\sim$0.00})& 0.896 $\pm$ 0.044 (\tcolB{+0.05})\\
\midrule

NV-WNS & 0.896 $\pm$ 0.073 (\tcolB{+0.05})& 0.861 $\pm$ 0.111 (\tcolB{+0.02}) & 0.896 $\pm$ 0.073  (\tcolB{+0.05})& 0.855 $\pm$ 0.056 (\tcolB{+0.01}) & \cellcolor{gray!15}\textbf{0.914 $\pm$ 0.068 (\tcolB{+0.07})} & 0.877 $\pm$ 0.087 (\tcolB{+0.03})&0.893 $\pm$ 0.042 (\tcolB{+0.05})\\

AA-W2V &0.893 $\pm$ 0.042 (\tcolB{+0.05})& 0.918 $\pm$ 0.068 (\tcolB{+0.07})& 0.858 $\pm$ 0.035 (\tcolB{+0.01})& 0.918 $\pm$ 0.068 (\tcolB{+0.07})& 0.896 $\pm$ 0.073 (\tcolB{+0.05})& 0.893 $\pm$ 0.042  (\tcolB{+0.05})& \cellcolor{gray!15}\textbf{0.922 $\pm$ 0.068 (\tcolB{+0.08})}\\

T-WNL & 0.871 $\pm$ 0.037 (\tcolB{+0.03})& \cellcolor{gray!15}\textbf{0.918 $\pm$ 0.068 (\tcolB{+0.07})}& \cellcolor{gray!15}\textbf{0.918 $\pm$ 0.068 (\tcolB{+0.07})}& 0.896 $\pm$ 0.073 (\tcolB{+0.05}) & 0.896 $\pm$ 0.073 (\tcolB{+0.05})& \cellcolor{gray!15}\textbf{0.918 $\pm$ 0.068 (\tcolB{+0.07})}& \cellcolor{gray!15}\textbf{0.918 $\pm$ 0.068 (\tcolB{+0.07})}\\

Combined DA &0.896 $\pm$ 0.073 (\tcolB{+0.05})& 0.871 $\pm$ 0.037 (\tcolB{+0.03})& 0.918 $\pm$ 0.068 (\tcolB{+0.07})& \cellcolor{gray!15}\textbf{0.952 $\pm$ 0.067 (\tcolB{+0.11})}& 0.918 $\pm$ 0.068 (\tcolB{+0.07})& 0.918 $\pm$ 0.068 (\tcolB{+0.07})& 0.948 $\pm$ 0.036 (\tcolB{+0.10})\\

\bottomrule
\end{tabular}
    }} \\*[0.4em]
    
    \subfloat[OPENCOSS (No Augmentation conflict-F1: 0.683 $\pm$ 0.131)\label{tab:open_full_report}] {

\resizebox{0.985\textwidth}{!}{
\begin{tabular}{l|rrrrrrr}
    \toprule
        &I   & II  & III  & I+II & I+III & II+III & I+II+III   \\
    \midrule
Shuffling & 0.588 $\pm$ 0.068 (\tcolR{- 0.09})& 0.357 $\pm$ 0.254 (\tcolR{- 0.32})& 0.500 $\pm$ 0.081 (\tcolR{- 0.18})& 0.348 $\pm$ 0.247  (\tcolR{- 0.33})& 0.377 $\pm$ 0.031 (\tcolR{- 0.30})& 0.330 $\pm$ 0.272 (\tcolR{- 0.35})& 0.539 $\pm$ 0.179 (\tcolR{- 0.14})\\

Back Translation & 0.266 $\pm$ 0.188 (\tcolR{- 0.41})& 0.388 $\pm$ 0.283  (\tcolR{- 0.31})& 0.133 $\pm$ 0.188 (\tcolR{- 0.55})& 0.320 $\pm$ 0.227 (\tcolR{- 0.36})& 0.000 $\pm$ 0.000 (\tcolR{- 0.68})& 0.466 $\pm$ 0.094 (\tcolR{- 0.38})& 0.300 $\pm$ 0.216 (\tcolR{- 0.38})\\

Paraphrasing &0.460 $\pm$ 0.235 (\tcolR{- 0.22})&0.527 $\pm$ 0.171 (\tcolR{- 0.15})& 0.285 $\pm$ 0.404 (\tcolR{- 0.39})&0.487 $\pm$ 0.083 (\tcolR{- 0.19})&0.410 $\pm$ 0.090 (\tcolR{- 0.06}) &0.476 $\pm$ 0.356 (\tcolR{- 0.06}) &0.416 $\pm$ 0.311 (\tcolR{- 0.27})\\
\midrule

NV-WNS & 0.361 $\pm$ 0.331 (\tcolR{- 0.32})&0.522 $\pm$ 0.109 (\tcolR{- 0.16})& 0.355 $\pm$ 0.273 (\tcolR{- 0.32}) & 0.206 $\pm$ 0.182 (\tcolR{- 0.47})& 0.194 $\pm$ 0.141 (\tcolR{- 0.48})& 0.428 $\pm$ 0.349 (\tcolR{- 0.25})& 0.261 $\pm$ 0.204 (\tcolR{- 0.42})\\

AA-W2V &0.644 $\pm$ 0.273 (\tcolR{- 0.03})&0.555 $\pm$ 0.415 (\tcolR{- 0.12})& 0.644 $\pm$ 0.273 (\tcolR{- 0.03})&0.583 $\pm$ 0.180 (\tcolR{- 0.10})& 0.669 $\pm$ 0.195 (\tcolR{- 0.01})& 0.455 $\pm$ 0.087 (\tcolR{- 0.22})& 0.507 $\pm$ 0.367 (\tcolR{- 0.17})\\

T-WNL &0.472 $\pm$ 0.335 (\tcolR{- 0.21})& 0.478 $\pm$ 0.083 (\tcolR{- 0.20})&0.490 $\pm$ 0.070  (\tcolR{- 0.19})& 0.433 $\pm$ 0.224 (\tcolR{- 0.25})& 0.609 $\pm$ 0.188 (\tcolR{- 0.07})&0.605 $\pm$ 0.149 (\tcolR{- 0.07}) & 0.555 $\pm$ 0.078 (\tcolR{- 0.12})\\

Combined DA & 0.565 $\pm$ 0.142 (\tcolR{- 0.11})& \cellcolor{gray!15}\textbf{0.693 $\pm$ 0.254 (\tcolB{+0.01})} & 0.634 $\pm$ 0.044 (\tcolR{- 0.04})& 0.500 $\pm$ 0.136 (\tcolR{- 0.18})& 0.666 $\pm$ 0.188 (\tcolR{- 0.01})& 0.511 $\pm$ 0.279 (\tcolR{- 0.17})& 0.563 $\pm$ 0.218 (\tcolR{- 0.12})\\

\bottomrule
\end{tabular}
    }} \\

\end{table*}

For WorldVista, Table~\ref{tab:w_v_full_report} provides a thorough evaluation of diverse DA configurations, employing both traditional and proposed DA techniques. The most notable performance improvement is achieved using the 'Shuffling' DA technique, yielding a conflict class F1-score of 90.8\% (+0.09) relative to the baseline, represented as an absolute value change. Further, for the WorldVista dataset, it can be observed that the most effective DA configuration for traditional DA techniques typically comprises Cases I+II+III. However, with the proposed DA techniques, there is no consistent pattern with respect to the effectiveness of DA configurations.

For the UAV dataset, as reported in Table~\ref{tab:uav_full_report}, we observe a significant performance increase using the `Combined DA' technique, specifically with the I+II+III configuration, resulting in a conflict class F1-score of 91.4\% and an absolute enhancement of +21. Additionally, the T-WNL technique demonstrates consistent efficacy across all cases for the UAV dataset, invariably leading to notable improvements in performance metrics. For the PURE dataset, as illustrated in Table~\ref{tab:pure_full_report}, the most substantial performance gain is achieved using the 'Combined DA' technique, which attains a 95.2\% F1-score under the I+II configuration. Interestingly, all of the proposed DA techniques consistently enhance performance for the PURE dataset, as demonstrated by the high positive shifts in performance scores. In contrast, traditional DA techniques seem to occasionally hinder performance values under certain configurations. 

In the case of the OPENCOSS dataset, we observe negligible enhancement in conflict class F1-scores, with the sole exception being Case II in the `Combined DA' technique. This lack of significant improvement could be attributed to the inherently limited number of conflict pairs in the original dataset. Furthermore, the application of DA techniques may have inadvertently disrupted the structure of sentence pairs within the OPENCOSS dataset, thereby leading to adverse learning effects on the transformer models.

\begin{figure*}[!ht]
    \centering
    \subfloat[Macro F1-score comparison \label{fig:doublechart_conflict}]{\includegraphics[width=0.475\textwidth]{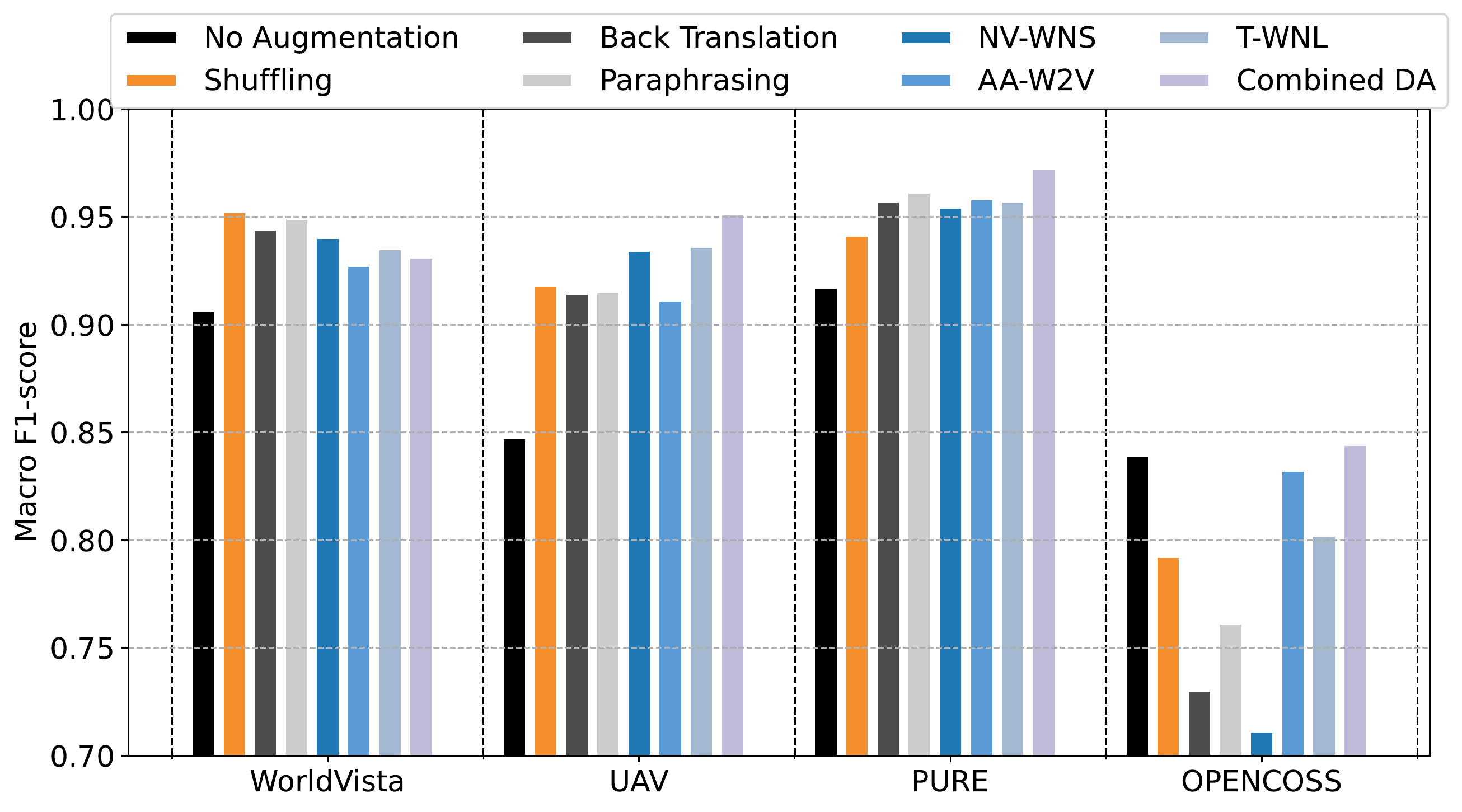}}
    \subfloat[Macro F1-score absolute improvement \label{fig:heatmap_conflict}]{\includegraphics[width=0.475\textwidth]{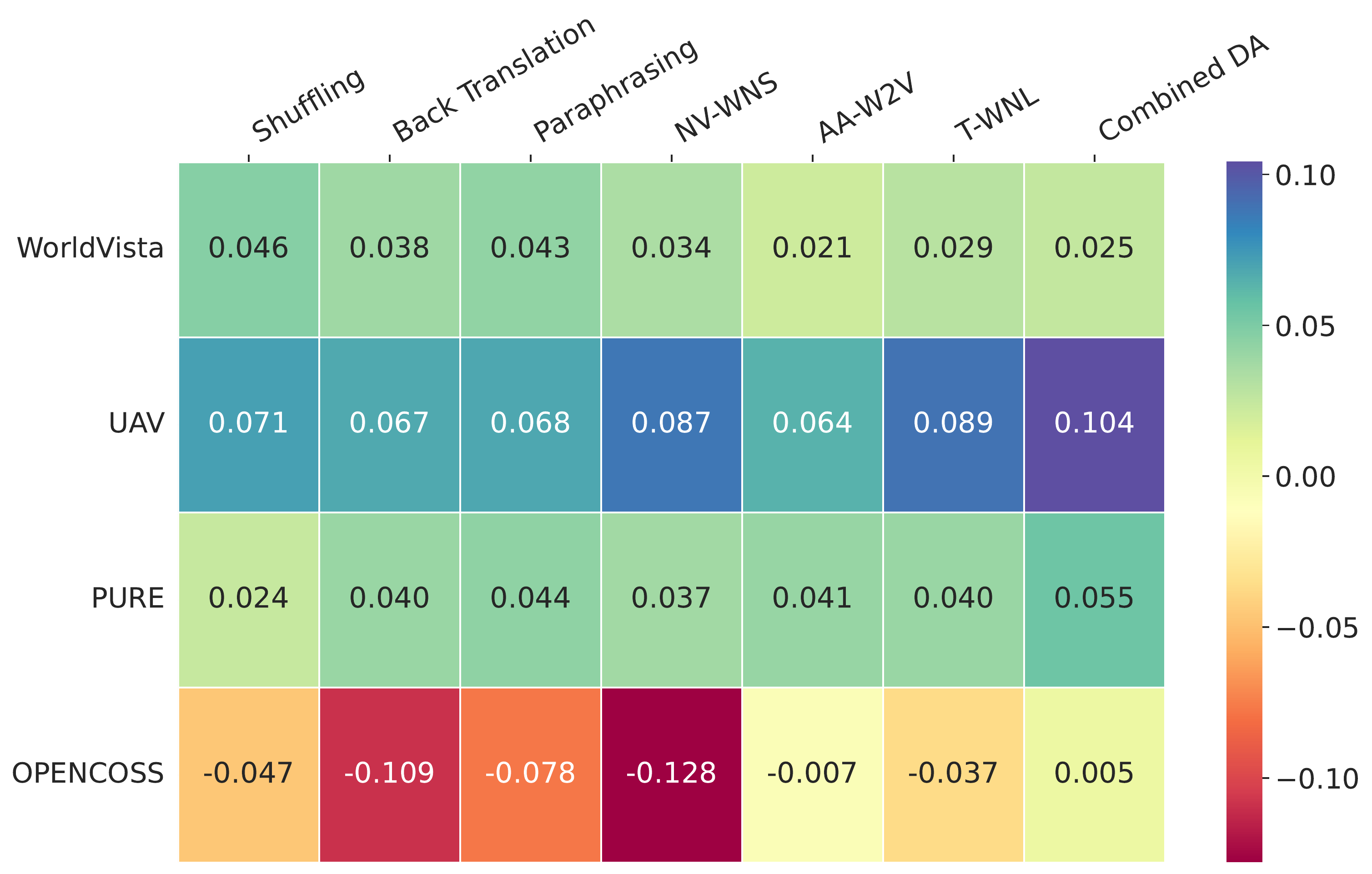}}
    \caption{Comparative analysis of macro F1-scores for the best performing DA configurations, as indicated in Table~\ref{tab:conflict_results}. This figure compares these scores with the `No Augmentation' baseline and includes a heat map to visually represent the absolute changes in macro F1-scores.}
    \label{fig:macro_f1_results_conflict}
\end{figure*}

Figure~\ref{fig:macro_f1_results_conflict} visualizes the influence of DA techniques on the macro averaged F1-scores for the binary classification task of neutral versus conflict in requirement pair datasets. Additionally, Figure~\ref{fig:doublechart_conflict} depicts the macro F1-scores for the best cases outlined in Table~\ref{tab:conflict_results} for the same datasets. Correspondingly, Figure~\ref{fig:heatmap_conflict} presents the absolute variations in the macro F1-scores. Here, OPENCOSS exhibits a negative trend, whereas WorldVista, UAV, and PURE display positive enhancements. 

Table~\ref{tab:duplicate_results} presents the evaluation outcomes for software platform duplicate post-classification datasets. In the context of the StackOverflow dataset, as delineated in Table~\ref{tab:stack_full_report}, the `Paraphrasing' DA technique shows the highest improvement, achieving a duplicate class F1-score of 68.6\% and an increment of +45 in absolute value. Overall, all DA techniques demonstrate significant enhancements for the StackOverflow dataset, substantially increasing the duplicate class F1-score. Table~\ref{tab:bug_full_report} illustrates the performance evaluation for the BugZilla dataset, where the 'Shuffling' technique exhibits the greatest improvement, attaining an F1-score of 80.4\%.
\begin{table*}[!ht]
\centering
\caption{Empirical analysis of DA techniques on duplicate class F1-score for \textit{software platform datasets}. The reported values are averaged over three folds and presented as ``mean ± standard deviation''.}
\label{tab:duplicate_results}
    \subfloat[StackOverflow (No Augmentation duplicate-F1: 0.227 $\pm$ 0.161)\label{tab:stack_full_report}]{
\resizebox{0.985\textwidth}{!}{
\begin{tabular}{l|rrrrrrr}
    \toprule
        &I   & II  & III  & I+II & I+III & II+III & I+II+III   \\
    \midrule
Shuffling & \cellcolor{gray!15}\textbf{0.644 $\pm$ 0.037 (\tcolB{+0.41})}& 0.534 $\pm$ 0.099 (\tcolB{+0.30})& 0.567 $\pm$ 0.062 (\tcolB{+0.34})& 0.582 $\pm$ 0.005  (\tcolB{+0.35})& 0.596 $\pm$ 0.060 (\tcolB{+0.36})& 0.479 $\pm$ 0.171 (\tcolB{+0.25})& 0.582 $\pm$ 0.071 (\tcolB{+0.35})\\

Back Translation & 0.608 $\pm$ 0.068  (\tcolB{+0.38})& \cellcolor{gray!15}\textbf{0.620 $\pm$ 0.054  (\tcolB{+0.39})}& 0.586 $\pm$ 0.104 (\tcolB{+0.35})& 0.568 $\pm$ 0.056  (\tcolB{+0.34})& 0.582 $\pm$ 0.035 (\tcolB{+0.35})& 0.615 $\pm$ 0.138 (\tcolB{+0.38})& 0.537 $\pm$ 0.061 (\tcolB{+0.31})\\

Paraphrasing &0.608 $\pm$ 0.044 (\tcolB{+0.38})& 0.625 $\pm$ 0.049 (\tcolB{+0.39})& 0.625 $\pm$ 0.049 (\tcolB{+0.39})& 0.681 $\pm$ 0.043 (\tcolB{+0.45})& 0.607 $\pm$ 0.068 (\tcolB{+0.38})& 0.617 $\pm$ 0.028 (\tcolB{+0.39})& \cellcolor{gray!15}\textbf{0.686 $\pm$ 0.062 (\tcolB{+0.45})}\\

\midrule

NV-WNS &0.591 $\pm$ 0.069 (\tcolB{+0.36})& \cellcolor{gray!15}\textbf{0.652 $\pm$ 0.038  (\tcolB{+0.42})} & 0.593 $\pm$ 0.034  (\tcolB{+0.36})& 0.562 $\pm$ 0.044 (\tcolB{+0.33})& 0.506 $\pm$ 0.087 (\tcolB{+0.27})& 0.503 $\pm$ 0.060 (\tcolB{+0.27})& 0.599 $\pm$ 0.110 (\tcolB{+0.37})\\

Combined DA &0.552 $\pm$ 0.102 (\tcolB{+0.32})& 0.589 $\pm$ 0.073 (\tcolB{+0.36})& 0.582 $\pm$ 0.097 (\tcolB{+0.35})& 0.606 $\pm$ 0.047 (\tcolB{+0.37})& 0.571 $\pm$ 0.087 (\tcolB{+0.34})& 0.563 $\pm$ 0.218 (\tcolB{+0.33})& \cellcolor{gray!15}\textbf{0.631 $\pm$ 0.082 (\tcolB{+0.40})} \\

\bottomrule
\end{tabular}
}
} \\*[0.4em]
\subfloat[BugZilla (No Augmentation duplicate-F1: 0.699 $\pm$ 0.034)\label{tab:bug_full_report}] {
\resizebox{0.985\textwidth}{!}{
\begin{tabular}{l|rrrrrrr}
    \toprule
        &I   & II  & III  & I+II & I+III & II+III & I+II+III   \\
    \midrule
Shuffling & 0.580 $\pm$ 0.313 (\tcolR{- 0.11})& 0.772 $\pm$ 0.049 (\tcolB{+0.07})& 0.705 $\pm$ 0.106 (\tcolR{$\sim$0.00})&  \cellcolor{gray!15}\textbf{0.804 $\pm$ 0.008 (\tcolB{+0.10})} & 0.769 $\pm$ 0.022 (\tcolB{+0.07})& 0.754 $\pm$ 0.050 (\tcolB{+0.05})& 0.784 $\pm$ 0.022 (\tcolB{+0.08})\\

Back Translation & 0.713 $\pm$ 0.049 (\tcolB{+0.01}) & 0.724 $\pm$ 0.035 (\tcolB{+0.02})& 0.735 $\pm$ 0.061 (\tcolB{+0.03})& 0.719 $\pm$ 0.081 (\tcolB{+0.02})& \cellcolor{gray!15}\textbf{0.756 $\pm$ 0.041 (\tcolB{+0.05})} & 0.741 $\pm$ 0.039 (\tcolB{+0.04})& 0.747 $\pm$ 0.052 (\tcolB{+0.04}) \\

Paraphrasing &0.750 $\pm$ 0.031 (\tcolB{+0.05})& 0.752 $\pm$ 0.023  (\tcolB{+0.05})& 0.732 $\pm$ 0.038  (\tcolB{+0.03})& 0.753 $\pm$ 0.083 (\tcolB{+0.05})& 0.764 $\pm$ 0.051 (\tcolB{+0.06})& 0.770 $\pm$ 0.037  (\tcolB{+0.07})& \cellcolor{gray!15}\textbf{0.776 $\pm$ 0.049 (\tcolB{+0.07})}\\

\midrule

NV-WNS &0.684 $\pm$ 0.051 (\tcolR{- 0.01})& \cellcolor{gray!15}\textbf{0.736 $\pm$ 0.075 (\tcolB{+0.03})}& 0.696 $\pm$ 0.091 (\tcolR{$\sim$0.33})& 0.633 $\pm$ 0.051 (\tcolR{- 0.06})& 0.698 $\pm$ 0.045 (\tcolR{$\sim$0.33})& 0.687 $\pm$ 0.099 (\tcolR{- 0.01})& 0.725 $\pm$ 0.055 (\tcolB{+0.02}) \\

Combined DA & 0.737 $\pm$ 0.026 (\tcolB{+0.03})& 0.739 $\pm$ 0.065 (\tcolB{+0.04})& \cellcolor{gray!15}\textbf{0.758 $\pm$ 0.028 (\tcolB{+0.05})}& 0.673 $\pm$ 0.057 (\tcolR{- 0.02})& 0.704 $\pm$ 0.086 (\tcolB{$\sim$0.00})& 0.678 $\pm$ 0.085 (\tcolR{- 0.02}) & 0.711 $\pm$ 0.001 (\tcolB{+0.01}) \\

\bottomrule
\end{tabular}
    }
    }
\end{table*}

\begin{figure*}[!ht]
    \centering
    \subfloat[Macro F1-score comparison\label{fig:macro_f1_duplicate}]{\includegraphics[width=0.425\textwidth]{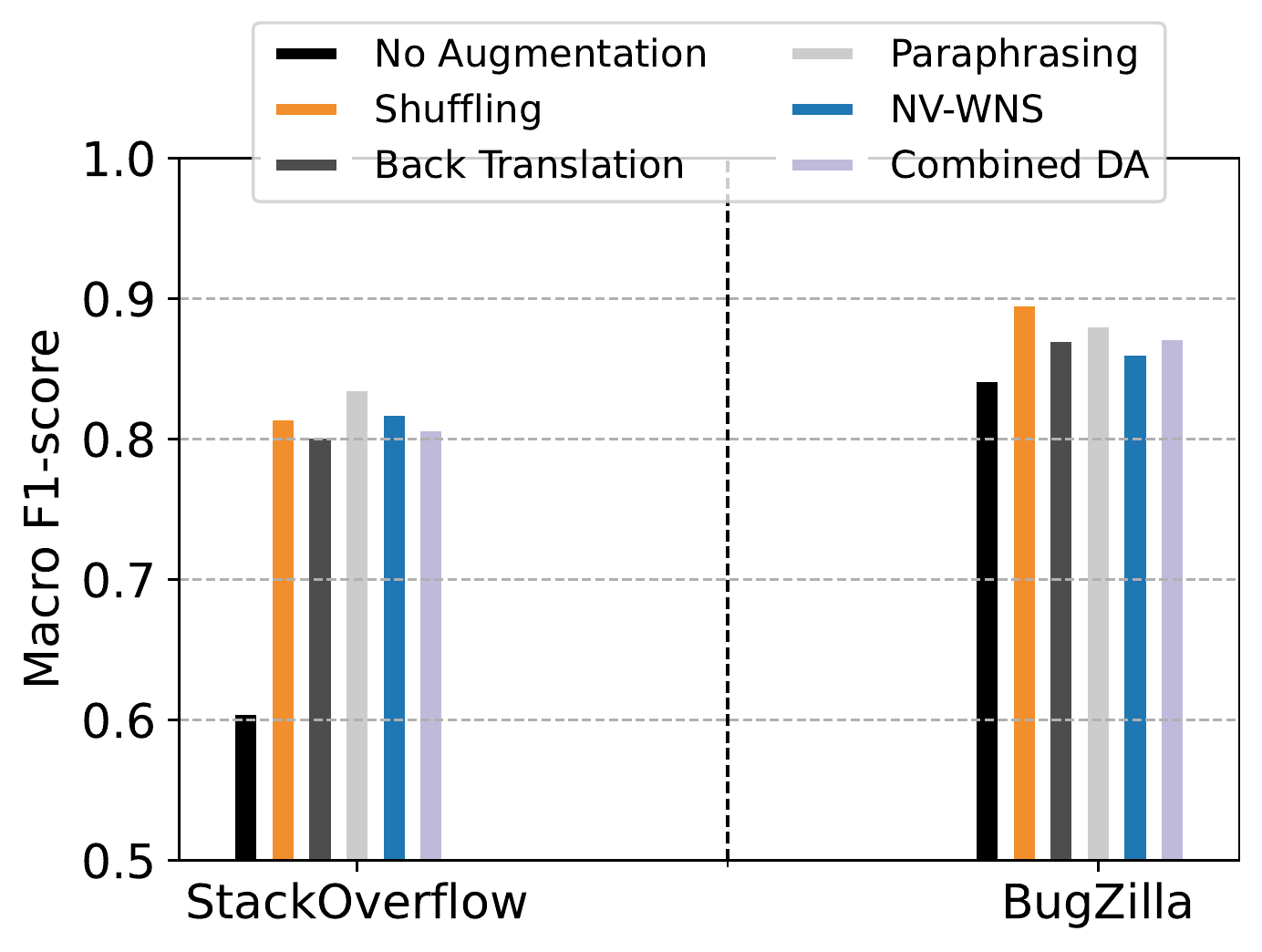}}\quad
    \subfloat[Macro F1-score absolute improvement\label{fig:heatmap_duplicate}]{\includegraphics[width=0.425\textwidth]{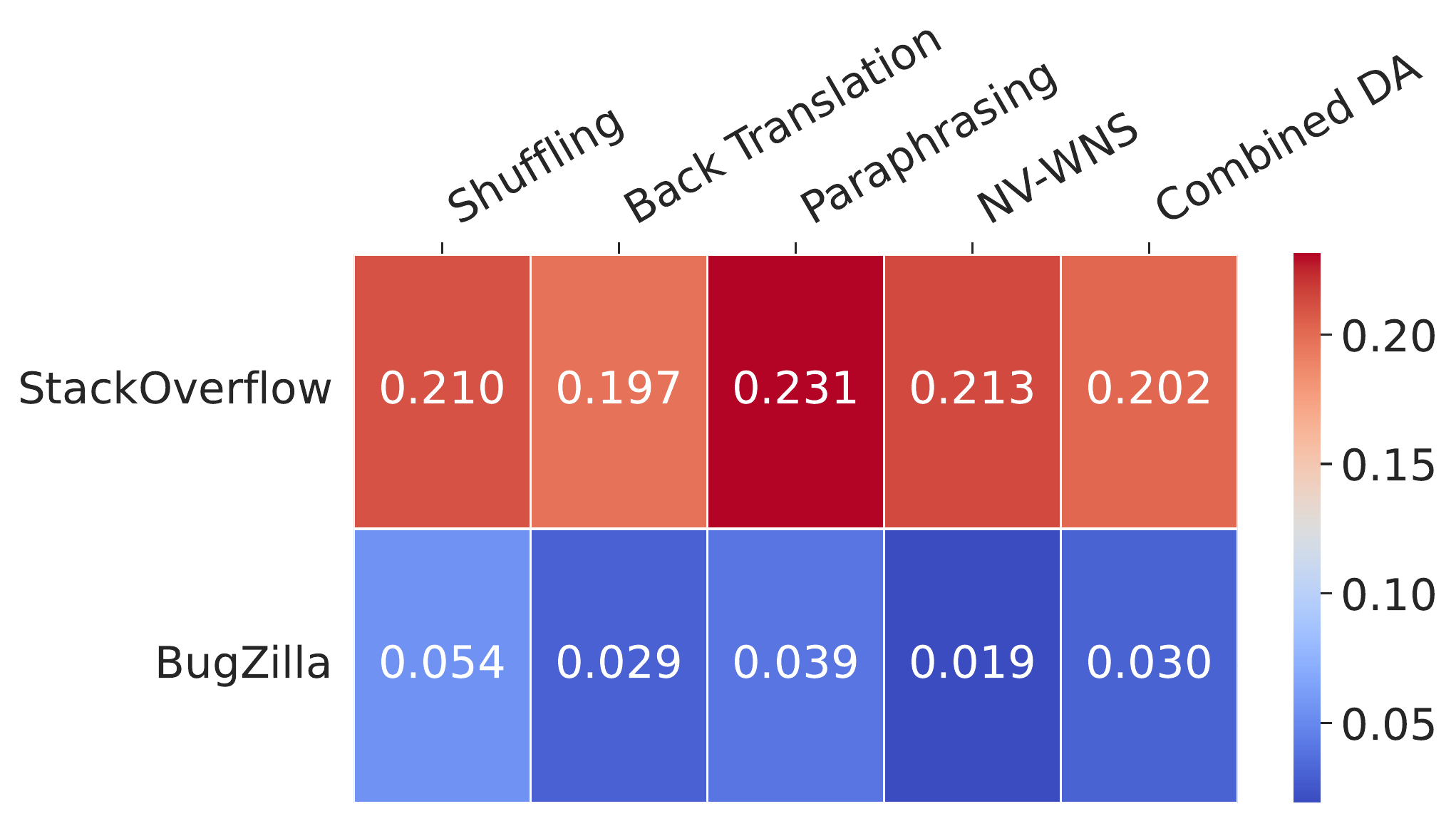}}
    \caption{Comparative analysis for software platform duplicate post datasets with macro F1-scores for the best performing DA configurations, as indicated in Table~\ref{tab:duplicate_results}.}
    \label{fig:macro_f1_duplicate_result}
\end{figure*}

Figure~\ref{fig:macro_f1_duplicate_result} illustrates the influence of DA techniques on the macro averaged F1-scores for the binary classification task of neutral versus duplicate in software platform duplicate-post datasets. Moreover, Figure~\ref{fig:macro_f1_duplicate} depicts the macro F1-scores for the best cases outlined in Table~\ref{tab:duplicate_results} for the same datasets. Similarly, Figure~\ref{fig:heatmap_duplicate} shows the absolute changes in macro F1-score. StackOverflow demonstrates a substantial performance increase, while BugZilla also exhibits performance improvement, however, to a lesser extent.

In order to create Table~\ref{tab:improvement}, the best DA methods, in tandem with their best configurations, are selected from Tables~\ref{tab:conflict_results} and ~\ref{tab:duplicate_results}. Subsequently, both absolute and relative enhancement averages are calculated. Furthermore, the table also highlights the maximum performance enhancement for each DA method, specified for Precision, Recall, and F1-score metrics.
\begin{table*}[!ht]
\centering
    \caption{Average and maximum performance improvement for all the DA techniques presented as absolute / relative percentage change.}
    \resizebox{\textwidth}{!}{
    \begin{tabular}{l|cc|cc|cc}
\toprule
\textbf{DA Techniques} & \multicolumn{2}{c}{\textbf{Macro Precision}} & \multicolumn{2}{c}{\textbf{Macro Recall}} & \multicolumn{2}{c}{\textbf{Macro F1-score}} \\
\midrule
 & Avg. & Max. & Avg. & Max. & Avg. & Max. \\
 \cmidrule(lr){2-3}\cmidrule(lr){4-5}\cmidrule(lr){6-7}
Shuffling & +0.044 / 4.586\% & +0.105 / 13.565\% & +0.079 / 11.851\%& +0.205 / 35.901\% & +0.079 / 11.170\%& +0.210 / 34.482\%\\
Back Translation & +0.046 / 5.529\% & +0.107 / 13.943\% & +0.075 / 11.346\%& +0.204 / 35.820\%& +0.072 / 10.241\% & +0.197 / 32.488\%\\

Paraphrasing & \cellcolor{gray!15}\textbf{+0.057 / 7.028\%} & \cellcolor{gray!15}\textbf{+0.161 / 20.808\%} & +0.081 / 12.149\%& +0.216 / 37.993\%& \cellcolor{gray!15}\textbf{+0.083 / 11.828\%} & \cellcolor{gray!15}\textbf{+0.231 / 38.056\%} \\

\midrule
NV-WNS & +0.046 / 5.532\% & +0.123 / 14.109\%& +0.076 / 12.749\% & +0.231 / 40.602\%& +0.076 / 10.857\%& +0.231/ 35.072\%\\
AA-W2V & +0.041 / 4.787\% & +0.123 / 14.143\%& +0.038 / 4.368\%& +0.058 / 6.708\%& +0.043 / 4.937\% & +0.064 / 7.644\% \\
T-WNL & +0.043 / 4.831\% & +0.095 / 10.947\% & +0.043 / 5.052\%& +0.083 / 10.024\%& +0.045 / 5.136\% & +0.077 / 9.104\% \\
Combined DA & +0.024 / 2.988\% & +0.096 / 1.958\% & \cellcolor{gray!15}\textbf{+0.098 / 14.318\%}& \cellcolor{gray!15}\textbf{+0.236 / 41.410\%}& +0.082 / 11.433\% & +0.202 / 33.294\%\\
\bottomrule
\end{tabular}

    }
    \label{tab:improvement}
\end{table*}
\subsection{Incremental Analysis} 
We conduct experiments to assess the impact of augmenting the number of instances within the conflict and duplicate classes on the performance outcomes for the WorldVista, Stack Overflow, and Bugzilla datasets. For this experiment, we select the most effective DA techniques from Tables~\ref{tab:conflict_results} and ~\ref{tab:duplicate_results} for the aforementioned datasets. Given that StackOverflow and BugZilla are open-source datasets, we were able to gather additional instances for the duplicate class.

\begin{figure*}[!ht]
     \centering
     \subfloat[WorldVista \label{fig:w_v_increment}]{\includegraphics[width=0.315\textwidth]{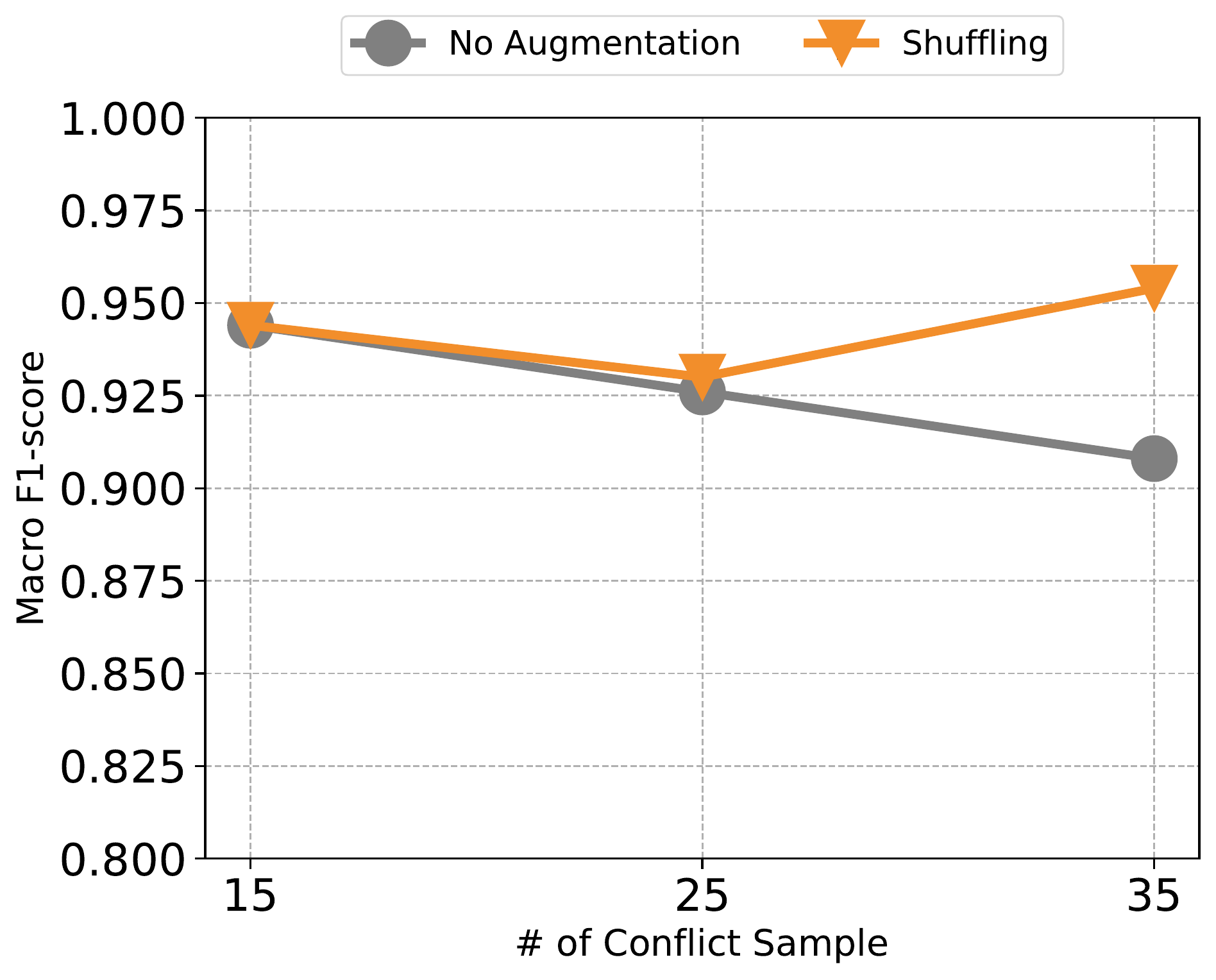}}
     \subfloat[StackOverflow \label{fig:stack_increment}]{\includegraphics[width=0.315\textwidth]{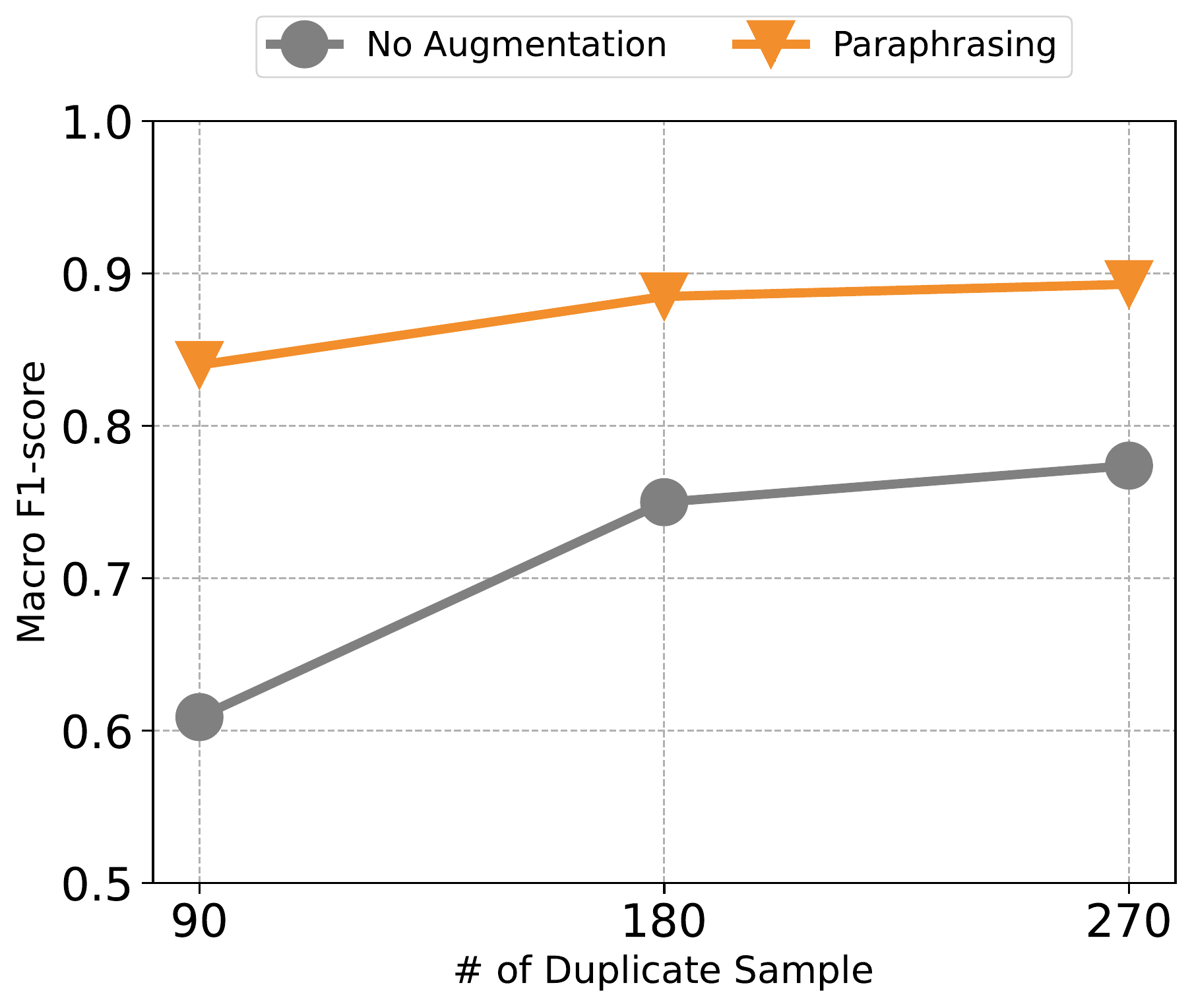}}
     \subfloat[BugZilla \label{fig:bug_increment}]{\includegraphics[width=0.315\textwidth]{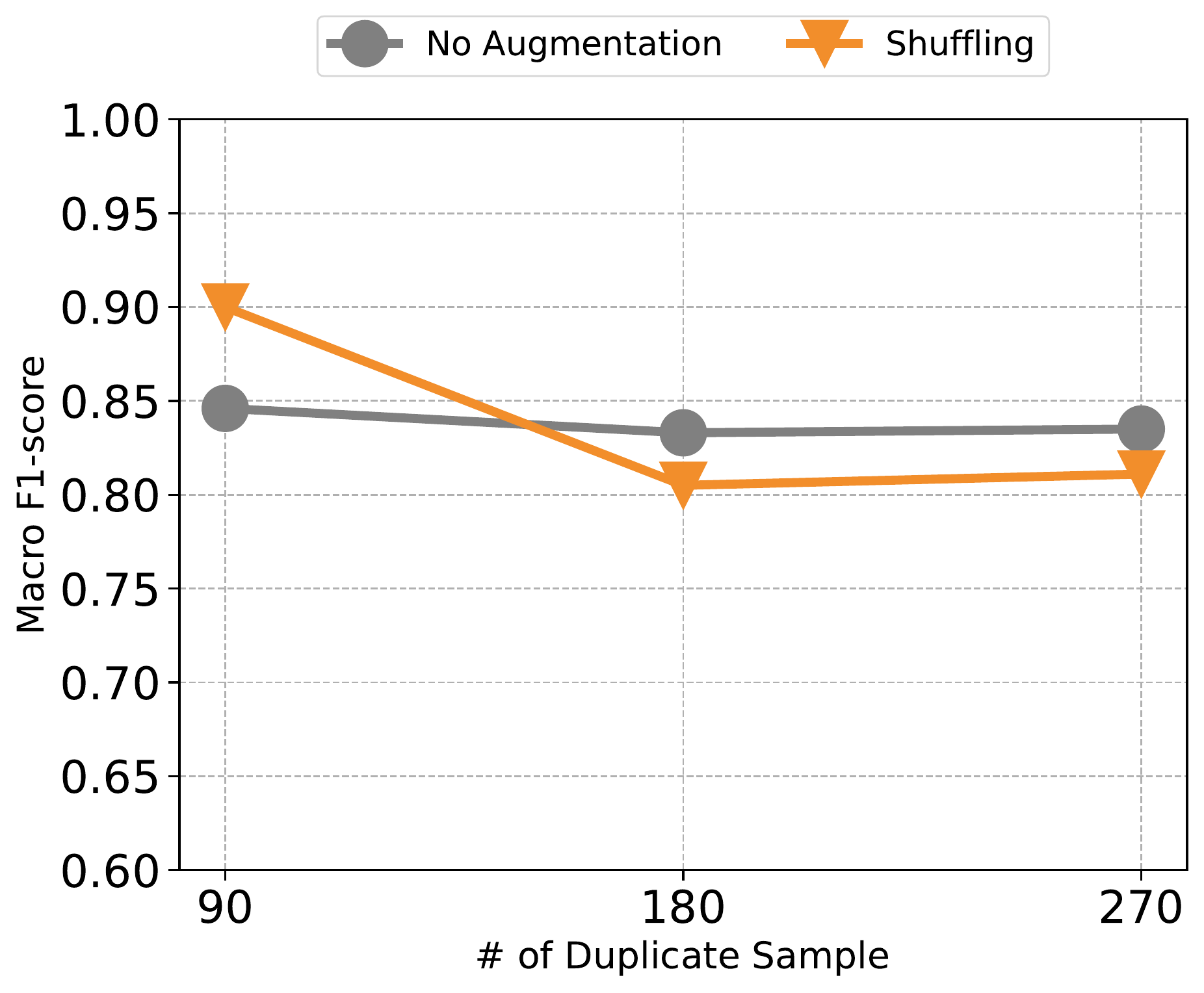}}
     \caption{Evaluating the performance of the most effective DA techniques for both conflict and duplicate classes, given various training set sizes.}
     \label{fig:increment_change}
\end{figure*}

In the case of the WorldVista dataset, as depicted in Figure~\ref{fig:w_v_increment}, we observe comparable performance outcomes for both the 'No Augmentation' baseline and DA techniques when applied to 15 and 25 conflict samples. However, a discernible improvement is noted when the sample size is increased to 35. The performance comparison between the 'No Augmentation' and 'Paraphrasing' techniques in relation to the macro F1-score for the Stack Overflow dataset is illustrated in Figure~\ref{fig:stack_increment}. A similar trend is identifiable in the performance specific to the duplicate class. As the quantity of duplicate class instances is increased from 180 to 270, the performance gap between the non-augmented and paraphrased instances diminishes. This suggests that a larger original sample size in training could potentially lead to a negative impact on the performance of DA techniques.

The performance of the duplicate class for the Bugzilla dataset, comparing the 'No Augmentation' and 'Shuffling' techniques, is shown in Figure~\ref{fig:bug_increment}. As hypothesized, the application of DA techniques yielded consistent results for smaller duplicate class instances. Yet, for larger duplicate class instances, it negatively affected model performance.

\section{Conclusion}\label{sec:conc}
Data augmentation techniques can generate synthetic input instances for domain-specific machine learning problems. In this study, data augmentation techniques are utilized in order to enhance the performance of the sentence pair classification tasks that are frequently encountered in the software engineering domain. We employ NLP-based techniques for data augmentation such as shuffling, back translation, paraphrasing, synonym substitution for nouns and verbs, target-lemma substitution, and synonym replacement for actors and actions using word embedding. Six different software pair text classification datasets are used in the experiments. We show that across all datasets, DA approaches significantly increase performance for conflict and duplicate classes. We further demonstrate that when the class proportions are relatively well-balanced, DA approaches have a detrimental effect on model performance. This work can be further extended in different directions.
First, training software-specific text embeddings can help find more relevant words for replacement in the data augmentation strategies. Also, we can study advanced DA techniques used in other domains to enhance the quality of augmented training data for sentence pair classification tasks. While we consider a large number of transformer models in our preliminary analysis, in future research, it can be explored which transformer models benefit more from data augmentation. Lastly, our analysis can be extended to sentence pair classification tasks seen in other domains.
  
\section*{Data Availability}
\url{https://gitfront.io/r/user-9946871/ii6eJFSh7oT4/DA-Sentence-Pairs/}


\bibliographystyle{ACM-Reference-Format}
\bibliography{Ref}

\end{document}